\documentclass{report}
\usepackage[ansinew]{inputenc}
\usepackage[]{epsfig}

\renewcommand{\abstract}[1]{{\footnotesize \noindent {\bf Abstract} #1 \\}}
\renewcommand{\author}[1]{\subsection*{#1}}
\newcommand{\address}[1]{\subsection*{\it#1}}

\newcommand{\Mib}{$\mbox{Milano } \beta \beta$}

\begin{document}

\chapter*{Low-Temperature Direct Dark Matter Searches}
\author{P.~C.~F.~Di~Stefano\footnote{Present address: IPNL, 4 rue Enrico Fermi, F-69622 Villeurbanne, France, distefano@ipnl.in2p3.fr}}
\address{Max-Planck-Institut für Physik, Föhringer Ring 6, \\ D-80805 Munich, Germany}

 \abstract{Small cryogenic detectors with efficient background
 rejection now best longer established and heavier direct
 dark matter searches.  This paper reviews the experiments, results
 and prospects.}

\section{Introduction}

Much evidence suggests a significant amount of non-baryonic dark
matter in the Universe, in the guise of putative Weakly
Interacting Massive Particles (WIMPs), such as the supersymmetric
neutralino. There are indications such particles may be present in
the Milky Way --- hence the motivation to detect them directly,
rather than products of their annihilation elsewhere.  These
various issues are discussed in Ref.~\cite{proc:IDM2000}.

Cryogenics, in fact,  played a part in starting the direct dark
matter searches. Low temperature super-conducting grains were
suggested as neutrino detectors in the early
1980s~\cite{art:Drukier}.  It was then pointed out that such
devices should also work to detect recoils caused by the elastic
scattering of WIMPs~\cite{art:Goodman}. Unfortunately,
super-conducting grains turned out to have several practical
problems. Indeed the first results from direct  searches were
obtained by detectors using earlier technologies, first ionization
in semi-conductors~\cite{art:ionisation_old}, then scintillation
detectors~\cite{art:NaI_1992}. Cryogenic detectors, in the guise
of calorimeters, have only recently begun to prove their worth for
dark matter searches.

\begin{figure*}
  \centering
        \mbox{\epsfig{file=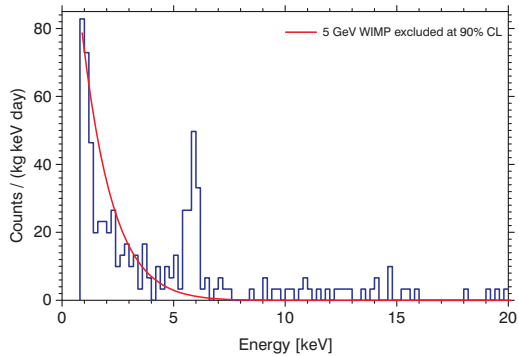,width=0.5\linewidth}}
  \caption{Expected $5 \mbox{ GeV/c}^2$ WIMP spectrum on sapphire (curve),
  and example of experimental background observed in the CRESST
  experiment~\cite{proc:CRESST_Lepton2001}. WIMP spectra are expected to have a
  quasi-exponential shape, of steepness decreasing with WIMP mass.}
  \label{fig:distefano_spectra}
\end{figure*}

Generally speaking, direct dark matter searches are challenging
because the expected exponentially decaying recoil energy spectrum
yields  rather small recoil energies (of the order of keV), and few
events above threshold, at most a few per day and per kilogram of
detector (Fig.~\ref{fig:distefano_spectra}).
At first glance therefore, detectors for WIMPs require  a high
mass, a low background and a low threshold. An additional
criterion is to have various target nuclei available to enable
cross-checks and to study the two possible types of interaction,
spin-dependent and spin-independent (cross section $\propto A^2$).
The first generation of standard calorimeters excels at the last
three of these requirements (Sec.~\ref{sec:distefano_std_cal}); hybrid calorimeters circumvent the first requirement
(Sec.~\ref{sec:distefano_hyb_cal}).

\section{Standard calorimeters}
\label{sec:distefano_std_cal}
\subsection{Thermal phonon detectors}
\begin{figure*}
  \centering
    \begin{tabular}{ccc}
       \mbox{\epsfig{file=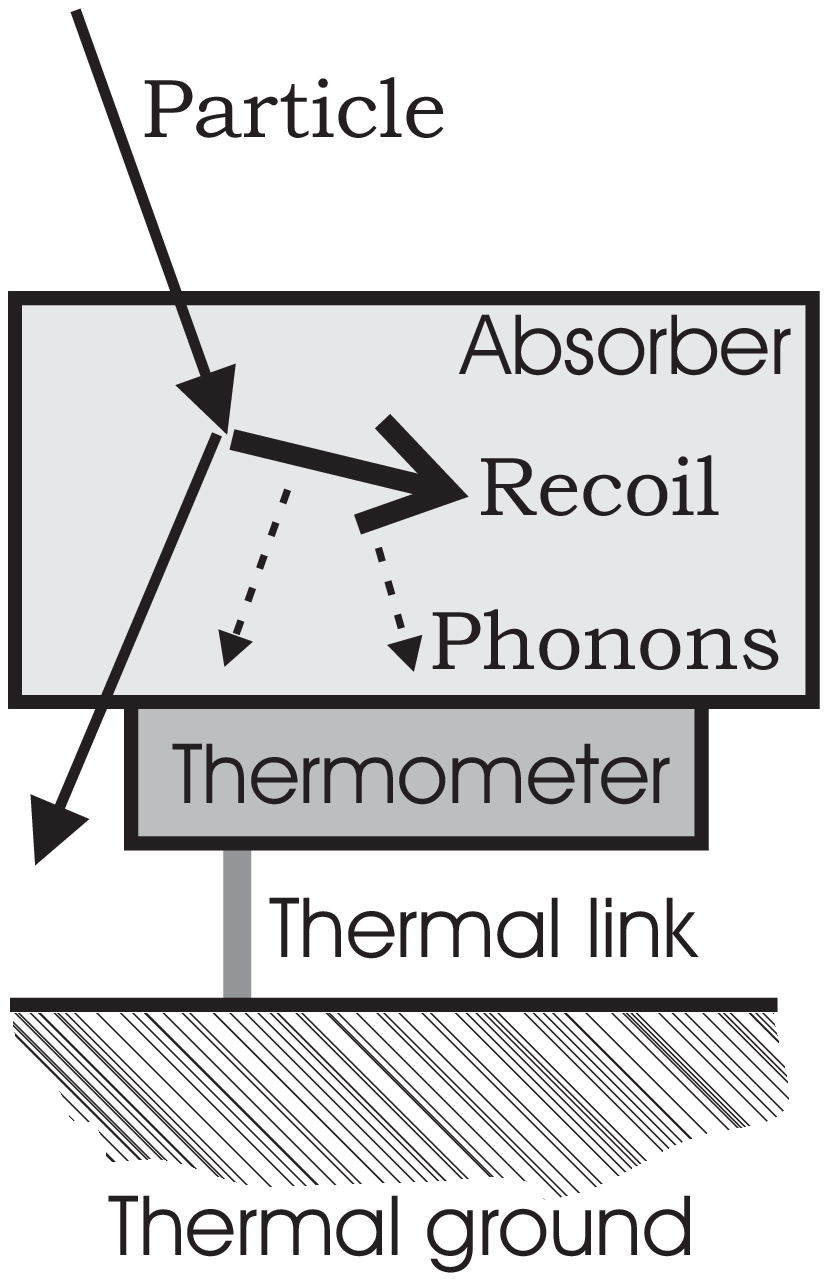,height=4cm}}  &
       \mbox{\epsfig{file=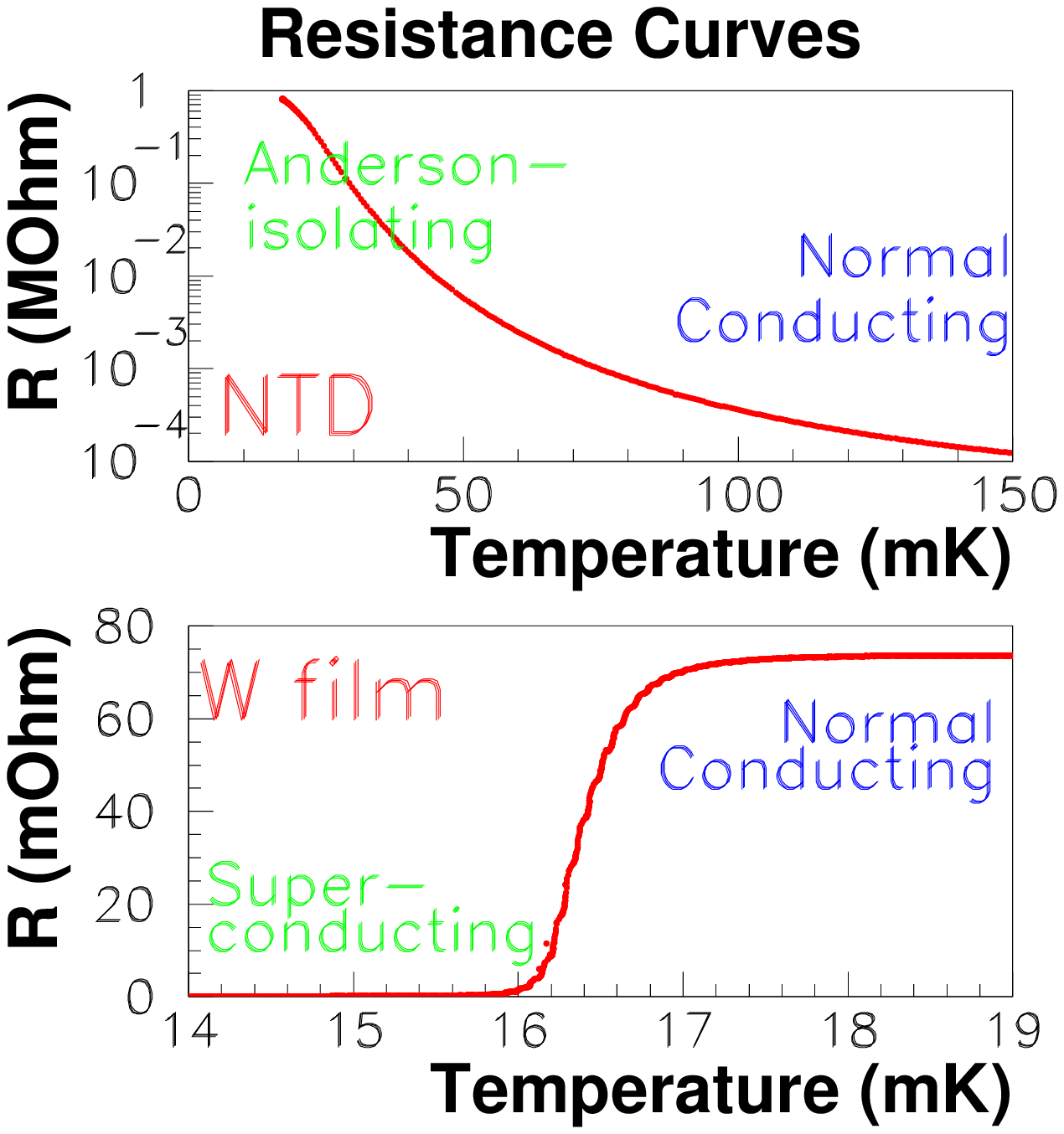,height=4cm}}   &
       \mbox{\epsfig{file=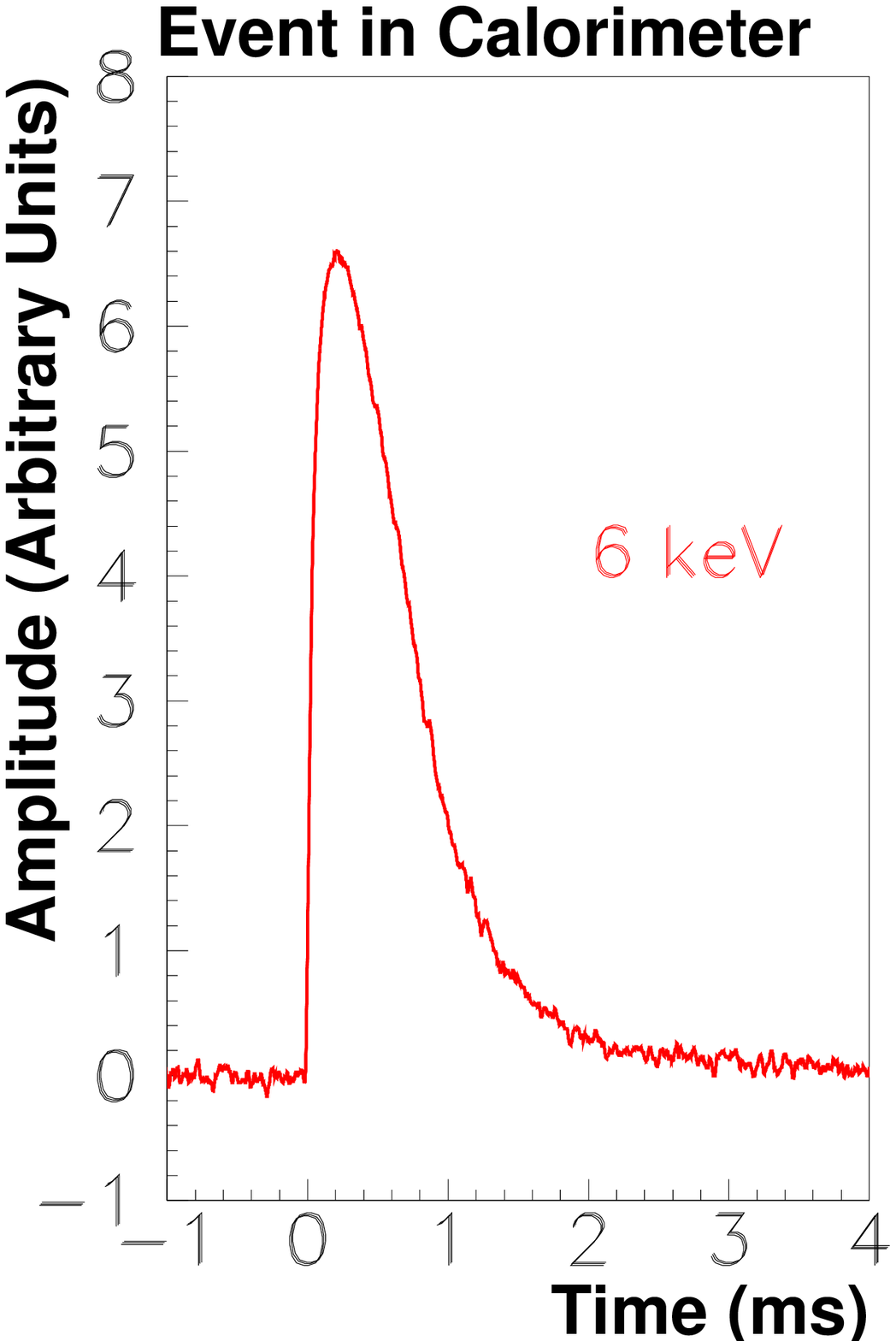,height=4cm}}
    \end{tabular}
  \caption{Left: principle of low temperature calorimetry.
  Center: examples of thermometers for calorimeters (NTD data courtesy A.~Juillard, CSNSM Orsay; other data MPI Munich).
  Note different R and T ranges.
  Right: 6~keV event in a calorimeter.}
  \label{fig:distefano_principle}
\end{figure*}

The original super-conducting grains are still being pursued by the
ORPHEUS experiment~\cite{proc:ORPHEUS_TAUP99}. However, the
cryogenic detectors most commonly used now in dark matter searches
are calorimeters (also referred to as bolometers in this field).  They are comprised of a main
crystal (the absorber), in which an incoming particle
scatters elastically off a nucleus, thereby releasing phonons which heat the
absorber up (Fig.~\ref{fig:distefano_principle}).  The variation
in temperature is read by a thermometer.  Because the temperature
rises are inversely proportional to the heat capacities involved,
and because specific heats of dielectrics decrease with
temperature, cryogenic conditions close to 10~mK are required to
obtain decent signal-to-noise ratios.  Most thermometers are
pieces of neutron-transmutation doped germanium (NTDs) with a
gradual change in resistance over a broad temperature range.

Experiments using this type of detector are listed in the first
parts of Tables~\ref{tab:distefano_exp}
and~\ref{tab:distefano_exp_2}. The Milano experiment is primarily
a search for  $\beta \beta$ decays, as were most of the early
direct dark matter searches~\cite{art:ionisation_old}. This
explains why the threshold of the experiment has not yet been
optimized.  Up to twenty 340~g detectors have been run
successfully  at a time; the fairly low background and the
presence of tellurium give this experiment the best
spin-independent limits of this first class of detector.

The Franco-Spanish {\it Rare Object Search Employing Bolometers
UndergrounD} (ROSEBUD) experiment has deployed various small
detectors made of sapphire and other materials. These detectors
show excellent thresholds, but suffer from an unfortunately high
background perhaps linked to the cryogenic equipment.  The Tokyo
experiment has tested up to 8 LiF units of 21~g each.  Despite a
high background due to the cosmic rays in the shallow site and to
microphonics, the experiment has good spin-dependent limits thanks
to the presence of fluoride. Improvements are expected as the
experiment moves to the deep Kamioka mine (2700~mwe).

\begin{figure*}[ht]
  \centering
    \begin{tabular}{cc}
        \mbox{\epsfig{file=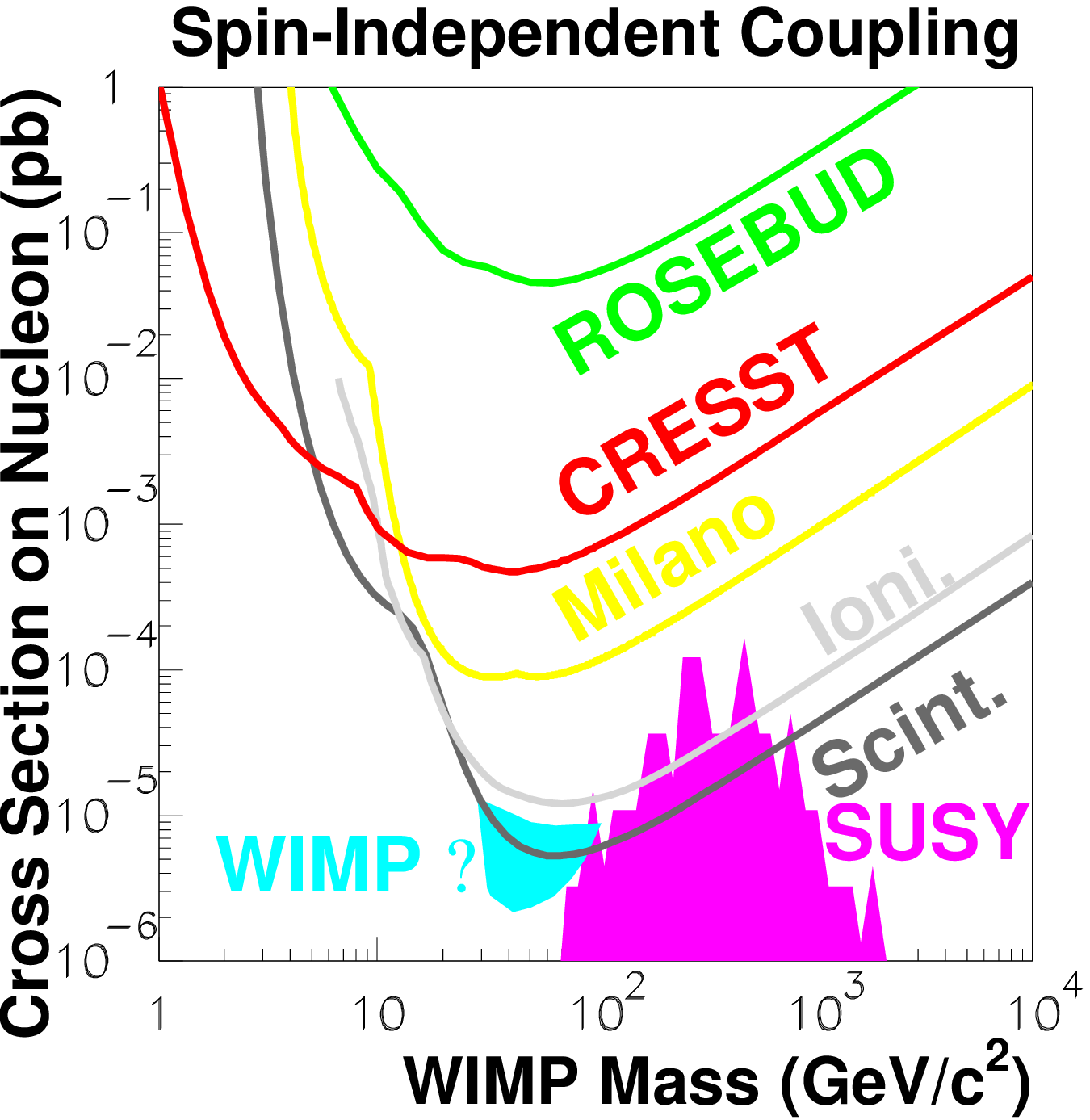,width=0.45\linewidth}} &
        \mbox{\epsfig{file=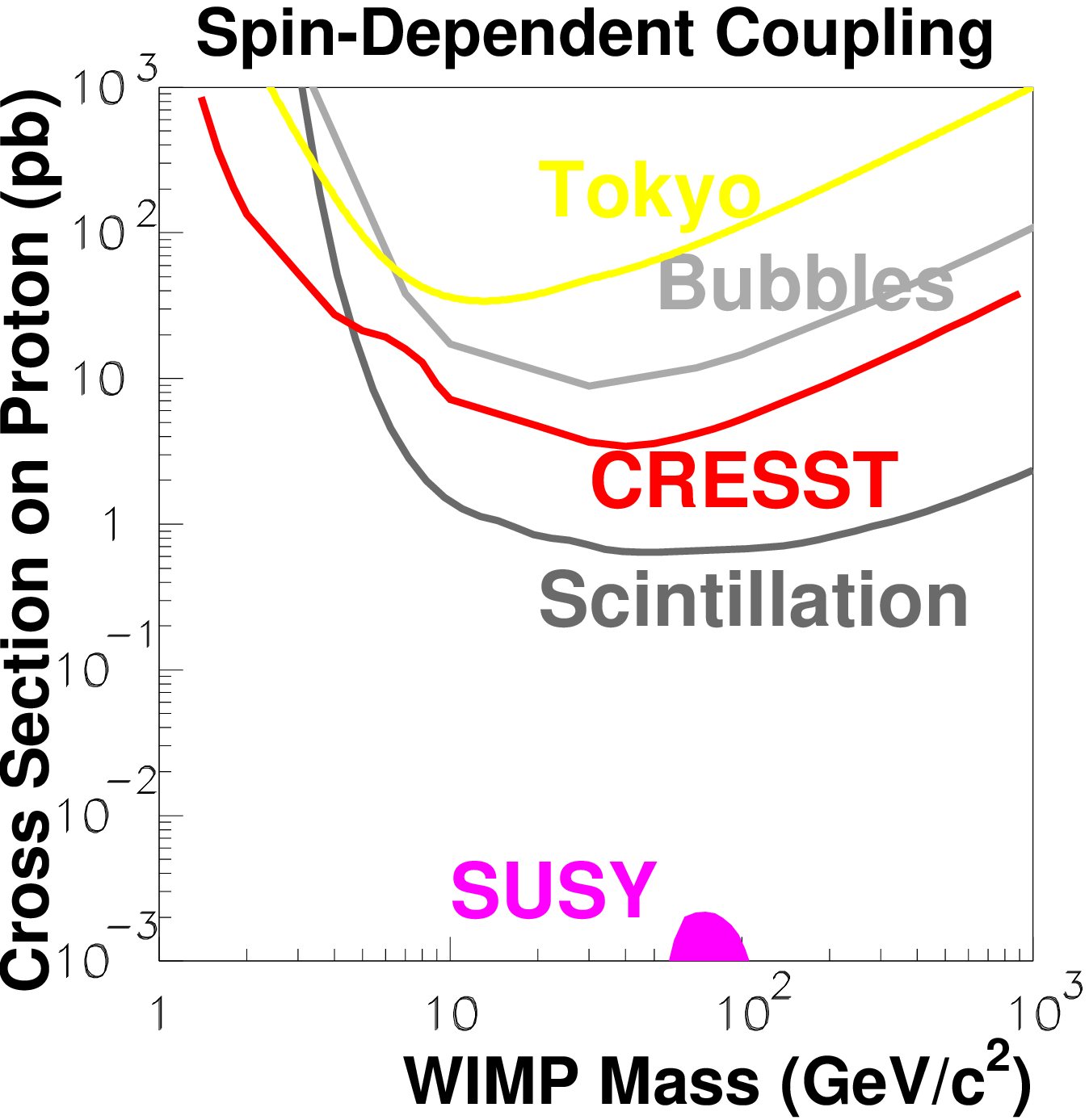,width=0.45\linewidth}}
    \end{tabular}
  \caption{Limits from standard calorimeters (90~\% CL), along with those from ionization~\cite{art:IGEX},
  scintillation~\cite{art:scintillation} and bubble detectors~\cite{art:bubbles}, approximate supersymmetric MSSM phase-space predictions (adapted from Ref.~\cite{pers:Gondolo,pers:Scopel}),
  and $3 \sigma$ WIMP claim~\cite{art:DAMA_2000}.
  The wide choice of target nuclei allows the study of spin-independent
  as well as spin-dependent interactions (although in the latter case, calorimeters,
  like most other direct
searches,
  are disadvantaged with regards to SUSY predictions by lack of coherent
  amplification factor $\sim A^2$ in cross-section).
  Standard calorimeters boast excellent thresholds enabling them to search for low mass WIMPs.}
  \label{fig:distefano_std_limits}
\end{figure*}

Experimental spectra are then compared with expected WIMP spectra
(Fig.~\ref{fig:distefano_spectra}) to set constraints on WIMP
cross-sections and masses (Fig.~\ref{fig:distefano_std_limits}).
Published limits are taken here at face value.  Though there is
convergence towards a standard set of astrophysical assumptions to
make these comparisons (most importantly a local WIMP density of
$0.3 \mbox{ GeV.cm}^{-3}$), some uncertainties remain in the
normalization of the spin-dependent limits~\cite{art:Tovey_2000},
and in the statistical methods to
use~\cite{proc:CRESST_Lepton2001,art:Green_2001}.
\begin{table*}[ht]
\caption{Current cryogenic direct detection experiments. Those in
deep sites tend to have lower backgrounds. Grains, thermal and
athermal phonon devices possess no active background
discrimination. A wide variety of target nuclei are available,
enabling studies of both spin-dependent and spin-independent
couplings. Ionization-phonon experiments have background
discrimination, and set best limits of any searches for
spin-independent coupling.} \label{tab:distefano_exp}
\begin{tabular}{|l|r|l|l|r|l|} \hline
  Name  & Depth  & Type  & Absorber  & Mass & Ref.  \\
  & (mwe) & & & (g) & \\
  \hline
  \hline
  ORPHEUS & 70 & Super-conducting grains & $\mbox{Sn}$ & $450$ & \cite{proc:ORPHEUS_TAUP99} \\
  \hline
  \Mib & 3500 & Thermal phonon & $\mbox{TeO}_2$ & $20 \times 340$ & \cite{art:Milano_2000} \\
  ROSEBUD & 2500 & Thermal phonon & $\mbox{Al}_2\mbox{O}_3 \mbox{, Ge}$ & 50 & \cite{astro-ph/0004292} \\
  Tokyo & 15 & Thermal phonon & $\mbox{LiF}$ & $8 \times 21$ & \cite{art:TokyoDM_PLB_1999} \\
  CRESST & 3500 & Athermal phonon & $\mbox{Al}_2\mbox{O}_3$ & $4 \times 262$ & \cite{proc:CRESST_Lepton2001} \\
  \hline
  CDMS & 20 & Ionization and phonon & $\mbox{Ge, Si}$ & $4 \times 165$ & \cite{art:CDMS_2000} \\
  EDELWEISS & 4500 & Ionization and phonon & $\mbox{Ge}$ & $320$ & \cite{art:EDELWEISS_PLB_2001} \\
  \hline
\end{tabular}
\end{table*}
\begin{table*}[ht]
\caption{Results from calorimetric experiments.  Threshold and
background are capital.  For ionization and phonon experiments
(second group), background is after rejection of electron recoils
but without any neutron background subtraction.  Energies are in true keV.}
\label{tab:distefano_exp_2}
\begin{tabular}{|l|r|r|l|} \hline
  Name & Threshold & Exposure & Approximate background  \\
   & (keV) & (kg.d) &  \\
  \hline
  \hline
  \Mib & 10 & 3.9 & 4 /d/kg/keV at threshold \\
  ROSEBUD & 0.5 & 0.4 & 15 /d/kg/keV over 25--50~keV \\
  Tokyo & 15 & 1.26 &  200 /d/kg/keV over 20--40~keV \\
  CRESST & 0.6 & 1.51 & 1 /d/kg/keV over 15--25~keV  \\
  \hline
  CDMS & 10 & 10.6 &  13 counts over 10--100~keV \\
  EDELWEISS & 30 & 4.53 &  0 counts over 30--200~keV\\
  \hline
\end{tabular}
\end{table*}

\subsection{Athermal phonon detectors}
The thermal phonon signal used in the detectors described so far has the drawback of vanishing quickly as absorber mass is
increased. As large absorber masses are of interest for direct
detection, it is judicious to collect and read high frequency
phonons in the thermometer before they thermalize in the
absorber~\cite{art:Proebst_1995}. This yields an amplification of
the pulses which is much less sensitive to absorber mass.
Thermometers used for this type of device are mainly thin films,
for instance using the sharp transition between the super-conducting
and normal-conducting states of tungsten (Fig.~\ref{fig:distefano_principle}).

The Anglo-German-Italian {\it Cryogenic Rare Event Search with
Super-conducting Thermometers} (CRESST) has developed and deployed
such devices using sapphire absorbers.  Up to four crystals of
262~g each have been run in the deep Gran Sasso site, in a
sophisticated cryogenic setup designed to minimize radioactive
contaminations~\cite{proc:CRESST_Lepton2001}.  These detectors
show excellent thresholds (600~eV) and low background
(Fig.~\ref{fig:distefano_spectra}).  CRESST is competitive for
light WIMPS, and now has the best calorimetric spin-dependent
limits thanks to the aluminum in its detectors.

\section{Hybrid calorimeters}
\label{sec:distefano_hyb_cal} Calorimeters can more than make up
for their small masses by rejecting background through a second
simultaneous measurement. This exploits the fact that  WIMPs (as
well as the neutron background) are expected to interact in the
absorber by creating nuclear recoils, whereas the photon and
electron backgrounds create electron recoils.

\subsection{Phonon-ionization detectors}
\begin{figure*}
  \centering
    \begin{tabular}{cc}
        \mbox{\epsfig{file=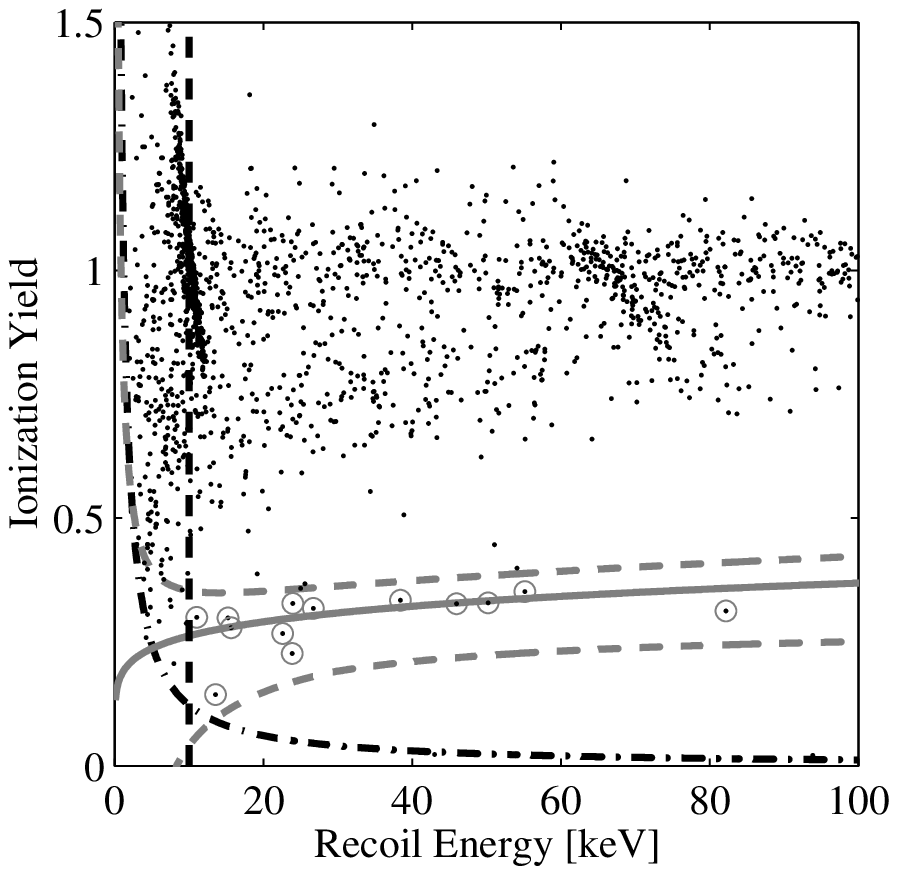,width=0.45\linewidth}} &
        \mbox{\epsfig{file=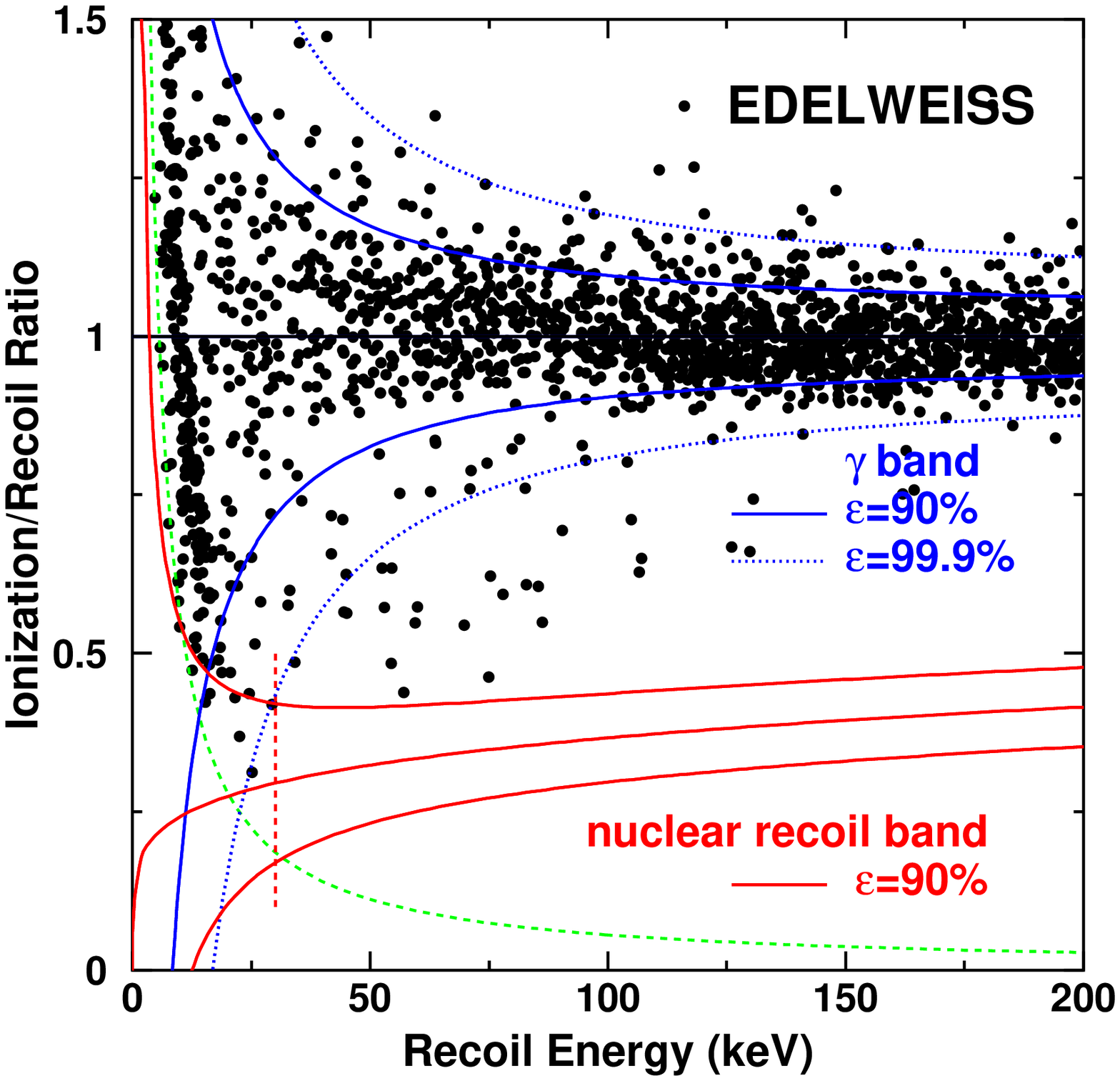,width=0.45\linewidth}}
    \end{tabular}
  \caption{Backgrounds from the CDMS~\cite{art:CDMS_2000} (left) and
  EDELWEISS~\cite{art:EDELWEISS_PLB_2001} (right)
  ionization-phonon experiments.
  For each event, ratio of ionization over recoil energy has been plotted as a function of recoil energy.
  Hyperbolae correspond to ionization threshold of hardware.
  Vertical segments are the recoil threshold used in analyses.
  Bands around a ratio of $\approx 0.3$
  are where 90~\% of nuclear recoils are expected, based on neutron calibrations.
  Most of the background has a ratio of $\approx 1$, corresponding to photon or electron background.
  Few or no events remain in nuclear recoil zone (13 events in 10.6~kg.d and 0 events in 4.53~kg.d respectively),
  demonstrating excellent background rejection.}
  \label{fig:distefano_cdms_edel}
\end{figure*}

\begin{figure*}
  \centering
    \begin{tabular}{cc}
        \mbox{\epsfig{file=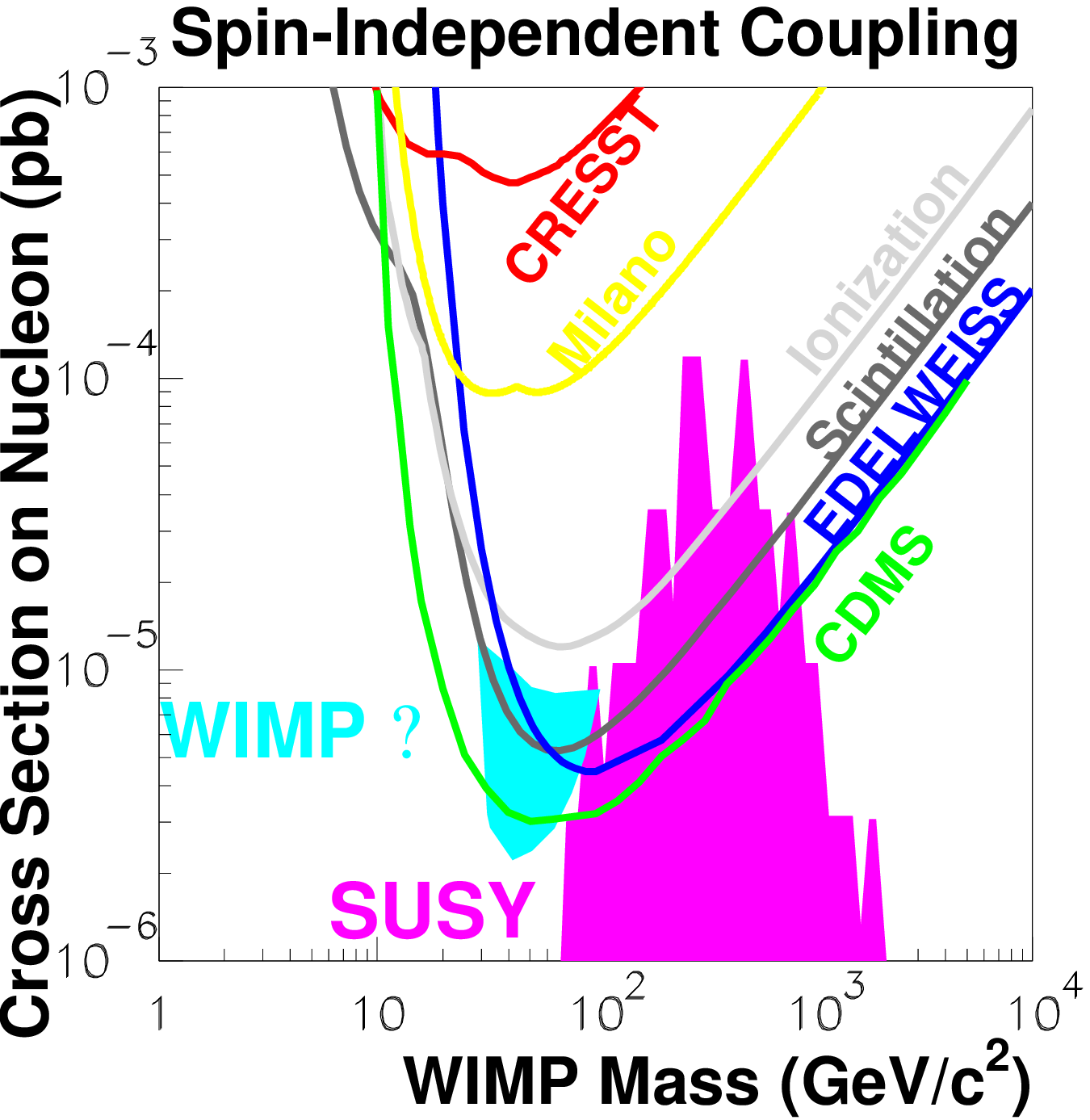,width=0.45\linewidth}} &
        \mbox{\epsfig{file=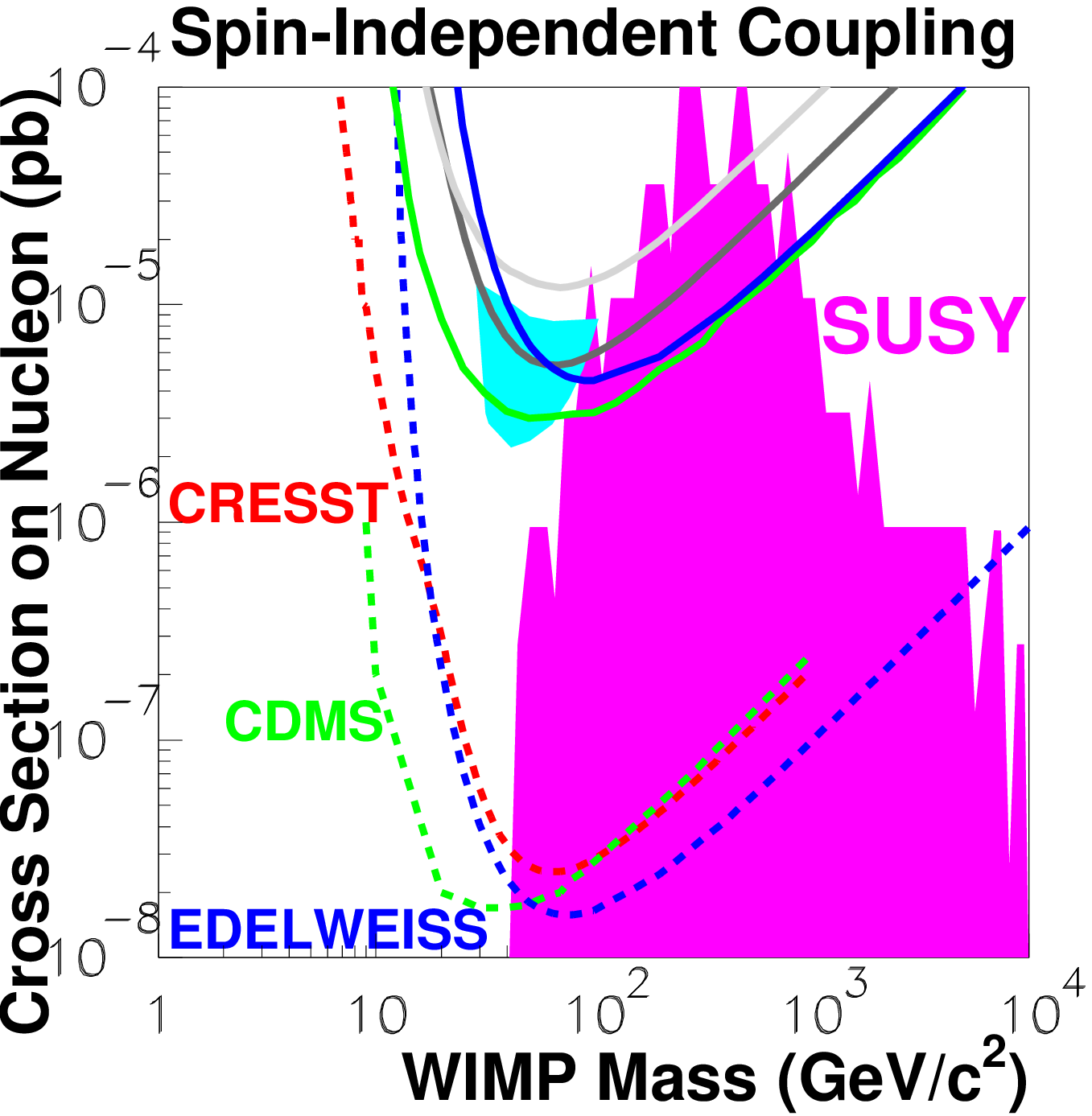,width=0.45\linewidth}}
    \end{tabular}
  \caption{Left: spin-independent limits from ionization-phonon calorimeters~\cite{art:CDMS_2000,art:EDELWEISS_PLB_2001}.
  Thanks to excellent background discrimination,
  these  detectors not only improve over their phonon-only counterparts,
  but surpass much heavier and more mature techniques.
  Paucity of spin in natural germanium makes for poor spin-dependent limits, not shown here.
  Right: spin-independent sensitivities expected from next-generation hybrid calorimetric experiments. Run times of $\approx 2$ years and mastery of neutron background are typically assumed.}
  \label{fig:distefano_hyb_limits}
\end{figure*}

The most developed technique so far uses calorimeters made out of
a semi-conductor.  The recoil imparted by a scattered particle
releases not only phonons but electron-hole pairs.  If the
surfaces of the absorber are implanted to make electrodes, a bias
voltage can be applied to drift the created charges to the
electrodes where they are counted, thereby producing a charge
signal in addition to the phonon signal.
Figure~\ref{fig:distefano_cdms_edel} shows that since electron
recoils create electron-hole pairs much more efficiently than
nuclear recoils do, this technique should allow rejection of
nearly 99.9~\% of the electron and photon background while
retaining about 90~\% of the nuclear recoils (from WIMPs and the
neutron background) and maintaining a decent threshold.
Teething problems with misidentified surface events which somewhat
limited the discrimination capacities~\cite{art:Edelweiss_2000_1}
now appear to have been contained thanks to various electrode
implantation schemes and position sensitive techniques.  The
American {\it Cryogenic Dark Matter Search} (CDMS) has deployed
several silicon and germanium devices. In a 1999 run, CDMS
obtained 13 nuclear recoils over 10.6~kg.d in three 165~g Ge
devices with a 10~keV threshold~\cite{art:CDMS_2000}.  This rate
would be compatible with the WIMP claimed by the DAMA NaI
scintillation experiment~\cite{art:DAMA_2000}. However, CDMS is in
a shallow site exposed to the neutron background created by cosmic
rays. Based on the number of multiple nuclear recoils observed,
and on the number of nuclear recoils previously observed in their
silicon detectors, CDMS reckons that most of the 13 events are in
fact neutron background, and subtracts some of them. This improves
the CDMS limit by a factor $\approx 2$, and greatly reduces the
compatibility with the DAMA result. More recently, the French {\it
Expérience pour DEtecter Les WIMPs En Site Souterrain} (EDELWEISS)
has operated a 320~g Ge device with a 30~keV threshold in a deep
site~\cite{art:EDELWEISS_PLB_2001}. The 4.53~kg.d of data show not
a single nuclear recoil.  The ensuing limit rules out the most
likely value of the DAMA NaI-1--4 WIMP with a 90~\% certainty.
Thus hybrid ionization-phonon calorimeters now have the best
spin-independent limits of all experiments
(Fig.~\ref{fig:distefano_hyb_limits}). 
Further results 
should be forthcoming
as CDMS moves to a deep site and EDELWEISS accumulates
more data.

\subsection{Scintillation-phonon detectors}
Their excellent results notwithstanding, ionization-phonon
detectors are by essence restricted to semi-conducting nuclei, in
practice Ge and Si.
For a wider choice of targets, the CRESST and ROSEBUD experiments
are developing a simultaneous measurement of scintillation and
phonons.
This is done by having a main calorimeter with an absorber made
out of an intrinsic scintillating material. In it, an incoming
particle releases not only phonons for the main phonon signal, but
also photons, which can escape the main calorimeter. With some
form of light collector, these photons can be detected in a
secondary calorimeter, giving the light signal. A 6~g proof of
principle experiment~\cite{art:Meunier1999} demonstrated
discrimination properties at least as good as those of
ionization-phonon detectors, with no complications arising from
surface events. $\mbox{CaWO}_4$, which should have an excellent
spin-independent cross section thanks to the tungsten, was used as
scintillator, though other materials also showed promise
($\mbox{BaF}_2$, BGO, and $\mbox{PbWO}_4$ where tested; other
tungstates and molybdates may function).
Efforts are under way to scale the device up  while maintaining a
low threshold.

\section{Prospects}
Hybrid ionization-phonon calorimeters now have the best
spin-independent limits.  Thanks to their wider choice of targets,
scintillation-phonon devices could prove an important complement
to them.  This is reflected in the ambitious next generation of
calorimetric experiments being planned in deep underground sites
(Tab.~\ref{tab:distefano_prospects}).
\begin{table*}
\caption{Next-generation calorimetric direct detection
experiments.    All will be in deep sites. CUORICINO, primarily a
$\beta \beta$ decay experiment, will have no background
discrimination, unlike the other experiments. The two
ionization-phonon ones, CDMS~II and EDELWEISS~II, should be
complemented by the scintillation-phonon one, CRESST~II, which
will provide a broader choice of target nuclei.  Partial results
could start coming within the next two years.  Experiments are
discussed in
Ref.~\cite{proc:IDM2000}.}\label{tab:distefano_prospects}
\begin{tabular}[b]{|l|r|l|l|r|l|} \hline
  Name & Depth & Type & Absorber & Mass \\
  & (mwe) &  &  & (g) \\
  \hline
  \hline
  CUORICINO & 3500 & Thermal phonon& $\mbox{TeO}_2$ & $56 \times 750$\\
  \hline
  CDMS II & 2100 & Ionization and phonon & $\mbox{Ge, Si}$ & $21 \times 250, 21 \times 100$ \\
  EDELWEISS II & 4500 & Ionization and phonon & $\mbox{Ge}$ & $21 \times 320$ \\
  CRESST II & 3500 & Scintillation and phonon & $\mbox{CaWO}_4$ & $33 \times 300$ \\
  \hline
\end{tabular}
\end{table*}
Apart from CUORICINO
--- a
scaled-up version of the \Mib\ experiment which will use standard
calorimeters, and whose assets for dark matter will be an
impressive total mass of 42~kg and the presence of tellurium (in
general good for spin-independent searches and in particular with
an atomic number very similar to that of iodine)
--- the experiments will all use hybrid calorimetric units of a
few hundred grams (ionization-phonon for CDMS~II and EDELWEISS~II,
scintillation-phonon for CRESST~II) deployed in arrays of up to
10~kg.
These experiments should have the sensitivity to start a thorough
exploration of supersymmetric phase space
(Fig.~\ref{fig:distefano_hyb_limits}).
Particularly in the spin-independent case, cryogenic detectors now
have a window of opportunity, since the earlier techniques of
ionization and scintillation which could hardly reject any
background require draconian radioactivity levels in order to improve, and other promising
background-insensitive techniques such as droplets or two-phase xenon are not yet fully
mature~\cite{art:bubbles,proc:ZEPLINIII_IDM2000}.

\section*{Acknowledgments}
The author thanks the organizers of the conference for their
invitation and a grant. D.~Akerib, M.~Altmann, G.~Chardin,
D.~Drain, E.~García, P.~de~Marcillac, O.~Martineau, K.~Miuchi, G.~Nollez,
S.~Pirro, K.~Pretzl, F.~Pröbst, B.~Sadoulet, R.~Schnee, S.~Scopel, W.~Seidel
and L.~Stodolsky provided helpful information and discussion. This
work was funded by European TMR Network for Cryogenic Detectors
ERB-FMRX-CT98-0167.

\end{document}